\documentclass[aip,rsi,reprint,superscriptaddress]{revtex4-1}
\usepackage{graphicx}
\usepackage{color}
\usepackage{array}
\usepackage{hyperref}
\hypersetup{colorlinks,citecolor=blue}
\begin{document}

\newcommand{\change}[1]{%
	{\color{blue}%
	\ensuremath{\clubsuit\!\triangleright}#1%
	\ensuremath{\triangleleft\!\clubsuit}}%
}

\title{Noise reduction of a Libbrecht--Hall style current driver}

\author{Christopher M. Seck}
\thanks{Christopher M. Seck and Paul J. Martin contributed equally to this work.}
\affiliation{Department of Physics and Astronomy, Northwestern University, Evanston, Illinois 60208-3112, USA}
\author{Paul J. Martin}
\thanks{Christopher M. Seck and Paul J. Martin contributed equally to this work.}
\author{Eryn C. Cook}
\affiliation{Department of Physics and Oregen Center for Optical, Molecular and Quantum Science, 1274 University of Oregon, Eugene, Oregon 97403-1274, USA}
\author{Brian C. Odom}
\email{b-odom@northwestern.edu}
\affiliation{Department of Physics and Astronomy, Northwestern University, Evanston, Illinois 60208-3112, USA}
\author{Daniel A. Steck}
\email{dan@steck.us}
\affiliation{Department of Physics and Oregen Center for Optical, Molecular and Quantum Science, 1274 University of Oregon, Eugene, Oregon 97403-1274, USA}

\begin{abstract}
The Libbrecht--Hall circuit is a well-known, low-noise current driver for narrow-linewidth diode lasers.  An important feature of the circuit is a current limit to protect the laser diode.  As the current approaches the maximum limit, however, the noise in the laser current increases dramatically.  This paper documents this behavior and explores simple circuit modifications to alleviate this issue.
\end{abstract}
\maketitle

\section{Introduction\label{sec:introduction}}

As their wavelength-coverage range continues to expand, diode lasers play an increasingly important role in experimental physics and related fields.  For many ultracold atomic and molecular experiments, the linewidth of the diode laser system must be significantly less than the natural linewidth of the addressed transition.  This requires the diode laser linewidth to be narrowed via an external cavity, forming an external-cavity diode laser (ECDL).  In this configuration, the linewidth of the ECDL is generally dominated by drive current noise.

For example, a typical AlGaAs laser diode undergoes $\sim$3~MHz/$\mu$A of frequency change with injection current.\cite{Wieman91}  To achieve a $\leq$1~MHz linewidth, the integrated current noise must be $\leq$300~nA.  In more precise applications, such as optical frequency standards, linewidths of $\ll$1~kHz are desired, necessitating a current drive noise $\ll$1~nA.  This level of current noise is probably unrealistic, but a low-noise current source lessens the demands on the servo system that is needed to narrow the laser's linewidth and stabilize it to an atomic or optical frequency reference.

\begin{figure*}[h]
\centering
\includegraphics[width=\textwidth]{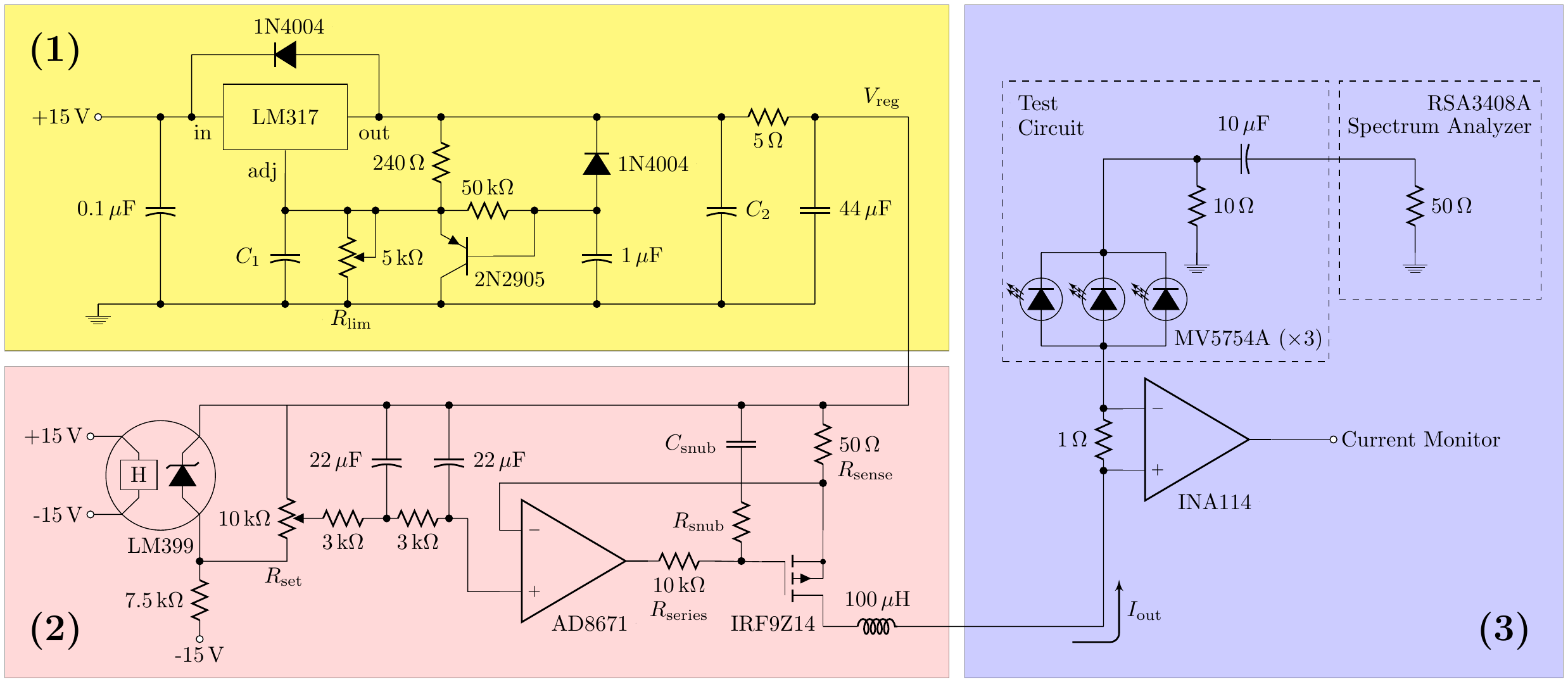}
\caption{Schematic of the Libbrecht--Hall-style test circuit: (1) voltage regulation, (2) active current-stabilization servo, (3) output current monitoring and noise measurement.  The elements of interest in this paper are the LM317 adjust pin bypass capacitor ($C_1$), the LM317 output bypass capacitor ($C_2$), and the snubber network ($C_\mathrm{snub}$ and $R_\mathrm{snub}$). Not included in the schematic are filtering networks on the power lines of the AD8671, where we use 10-$\Omega$ series resistors with 100-$\mu\mathrm{F}$ tantalum and 0.1-$\mu\mathrm{F}$ ceramic bypass capacitors.
\label{fig:schematic}}
\end{figure*}

The Libbrecht--Hall (LH) circuit, originally published in 1993, has been one of the primary circuits used for low-noise current drivers for laser diodes.\cite{Libbrecht93,Erickson08,Troxel11}  An implementation of the circuit (Fig.~\ref{fig:schematic}) contains several distinct subsections: (1) voltage regulation, (2) active current-stabilization servo, and (3) output current monitoring.  The LH circuit also contains a current-modulation section.  Since this functionality is not important for the present discussion, this section has been omitted from Fig.~\ref{fig:schematic}.

Section (1), the slow-turn-on, adjustable voltage source, centers on an LM317 voltage regulator that provides additional line regulation and defines a maximum current $I_\mathrm{max}$ for the circuit, excluding contributions from the current modulation section.  This maximum current is defined by the output voltage of the regulator and the voltage drops across all components between its output and ground, mainly the sense resistor $R_\mathrm{sense}$ and the laser diode.

Section (2), the current-stabilization feedback section of the circuit, uses an AD8671 op-amp in combination with an IRF9Z14 metal-oxide-semiconductor field-effect transistor (MOSFET) as a standard high-current source (see Fig.~4.12.A in \textcite{Horowitz}).  A resistor $R_\mathrm{series}$ in series with the op-amp output and a snubber network, comprising capacitor $C_\mathrm{snub}$ and resistor $R_\mathrm{snub}$, both act to stabilize the circuit at high frequencies.

In Section (3), an INA114 intrumentation amplifier measures the voltage across a $1~\Omega$ sense resistor to monitor the output current.  Ideally, this resistor is placed after the current-modulation junction to include its contribution.

The inclusion of the LM317 voltage regulator as the current limit in this circuit is clever: it effectively controls the power-supply voltage available to drive the laser, protecting it from inadvertent exposure to damagingly high currents.  However, interactions between the regulator, the MOSFET, and the AD8671 op-amp turn out to be the major contributors to the current noise of the circuit as $I\rightarrow I_\mathrm{max}$.  This means that the circuit cannot be used for low-noise current near $I_\mathrm{max}$, reducing the utility of the circuit, or even worse, unsuspecting users may be injecting more noise into their laser system than they realize by operating near $I_\mathrm{max}$.  The purpose of this work is to document these noise issues and to explore how they can be mitigated via simple component changes to a greater extent than in previous work.\cite{Erickson08,Troxel11}  The rest of this work has been organized into the following sections: (\ref{sec:vreg}.) voltage regulation problems; (\ref{sec:snub}.) op-amp feedback-loop stabilization; and (\ref{sec:test}.) adjustments of component values and measurement results.

\section{Voltage Regulation Problem\label{sec:vreg}}

The current-dependent inductive output impedance of the LM317 regulator is a commonly neglected issue with 3-terminal voltage regulators.\cite{Dietz}  Coupled with an output capacitance, this can produce a noise peak corresponding to the LC resonance.  The op-amp can only correct this to a certain extent.  However, the LM317 is not the \textit{primary} cause of the noise we observe in Fig.~\ref{fig:noise}(a): as current increases, the regulator's output inductance decreases,\cite{Dietz} pushing the resonant frequency higher instead of lower as we observe.

The LM317 noise can be reduced by increasing the values of $C_1$ and $C_2$ on the adjust pin and output of the regulator, respectively.\cite{Dietz}  In particular, as $C_2$ increases, the LC-resonance frequency decreases, and the op-amp can more readily prevent contamination of the output current due to the higher op-amp gain at low frequencies.

\section{Feedback-Loop Stabilization\label{sec:snub}}

The AD8671 stabilizes the output current of the IRF9Z14 p-channel MOSFET by controlling its gate voltage $V_\mathrm{G}$ so the source voltage $V_\mathrm{S}$ follows
the set point $V_\mathrm{set}$.  Because of the finite gain and high-frequency rolloff of the op-amp, its output impedance is effectively inductive (see Sec.~4.4.2.A in \textcite{Horowitz}).  This inductance, along with the MOSFET's gate capacitance, can produce a destabilizing resonance.

As noted in Refs.~\onlinecite{Libbrecht93,Erickson08,Troxel11}, external compensation components can stabilize an op-amp driving a capacitive load.  A series resistor $R_\mathrm{series}$ on the op-amp's output damps the LC resonance, and a snubber network, comprising $C_\mathrm{snub}$ and $R_\mathrm{snub}$, reduces the bandwidth of the current source.\cite{IntersilApp,MicrochipApp}

This network enables stable operation of the current driver, but does not completely solve the issue.  As $I \rightarrow I_\mathrm{max}$, the resonance behavior reemerges, although in a less dramatic fashion.  An increase in the MOSFET output current $I_\mathrm{SD}$ increases the source-gate voltage $V_\mathrm{SG}$ and decreases the source-drain voltage $V_\mathrm{SD}$.  Near $I_\mathrm{max}$, $V_\mathrm{SD}$ becomes small, and the transconductance $g_\mathrm{m}$ of the MOSFET decreases.  This effectively reduces the open-loop gain and increases the op-amp's output inductance.  We therefore expect the changing inductance to shift the noise \textit{downward} in frequency.  Note that a smaller $V_\mathrm{SD}$ implies a larger MOSFET gate capacitance, consistent with the  decrease in peak frequency. However, this is a smaller effect.  Additionally, the $Q$~factor of an RLC filter increases with the inductance.  We observe in Fig.~\ref{fig:noise}(a) a noise peak that both narrows and decreases in frequency as $I \rightarrow I_\mathrm{max}$.  This peak is consistent with the 12-kHz bump mentioned by LH\cite{Libbrecht93}, which they estimate contributes 16~nA of current noise.  Note that the data presented by LH\cite{Libbrecht93} was taken while operating likely at $<$$0.90\,I_\mathrm{max}$ for their test circuit, and the onset of the observed noise documented in this work begins when the set current is $>$$0.90\,I_\mathrm{max}$.

The inductor after the MOSFET also exacerbates the feedback instability.  The inductor's voltage drop increases at high frequencies, further reducing the MOSFET's high-frequency transconductance.  The inductor can also couple with the MOSFET's capacitance $C_\mathrm{SD}$ to create a low-impedance path for noise.  The current noise can be improved by bypassing the inductor with a short or resistor.\cite{Sauer}  However, this solution eliminates a desirable feature of the original LH design, allowing the possibility for an RF-modulation signal to destabilize the circuit.

Another simple solution to the noisy behavior near $I_\mathrm{max}$ could be to add a resistor in series with $R_\mathrm{set}$, limiting the output current maximum to less than that provided by $V_\mathrm{reg}$.  But this is equivalent to operating away from $I_\mathrm{max}$, and either prevents operation near the laser diode's true maximum current or undermines the protection provided by relying on $V_\mathrm{reg}$ to limit $I$ to $I_\mathrm{max}$.

In summary, the interaction of the op-amp and MOSFET introduces instability as $I \rightarrow I_\mathrm{max}$.  If the destabilization overlaps in frequency with the noise of the LM317 described in Sec.~\ref{sec:vreg}, then more noise is added to the output current.

Some simple alterations to the circuit address these effects.  First, we can increase $C_2$ to reduce the frequency of the LM317 noise peak.  The op-amp has higher low-frequency gain and is thus better able to handle power supply fluctuations at low frequencies.  Second, we can lower the roll-off frequency of the snubber network.  This further suppresses the effect of the op-amp/MOSFET resonance at the expense of op-amp bandwidth, so the snubber frequency should not be reduced excessively.  We cannot remove the inherent noisy behavior of the feedback loop because $g_\mathrm{m} \rightarrow 0$ as $I \rightarrow I_\mathrm{max}$.  However, we can improve how close we can get to $I_\mathrm{max}$ before the noise appears.

\section{Testing and adjustments\label{sec:test}}

For our op-amp and MOSFET, we have chosen components already discussed in Ref.~\onlinecite{Erickson08}.  The VP0106 FET in the original LH paper is replaced with an IRF9Z14 MOSFET due to higher current drive capacity (higher maximum drain current and power dissipation, and lower on-state resistance) while maintaining similar dynamic characteristics.  The AD8671 current-feedback op-amp offers slightly lower current noise than the original design's LT1028, but higher voltage noise. 

We use three LEDs (Fairchild MV5754A) connected via a 2-m shielded cable as a dummy load and record current-noise data with a Tektronix RSA3408A spectrum analyzer.  A capacitor blocks the dc signal and a $10$-$\Omega$ resistor converts current fluctuations to voltages that are monitored by the analyzer.  This method gives a higher noise floor than observed in Ref.~\onlinecite{Troxel11}, but enables a larger bandwidth.  We average 500 traces, plot the current spectral density, and quantify the noise by integrating the signal and subtracting the integrated baseline, obtained with the testing circuit disconnected from the switched-off current driver but still attached to the spectrum analyzer.  We measure the standard error of background noise to be less than 0.5~nA and assume a similar variation in the signal.

\begin{figure*}[h]
\centering
\includegraphics[width=\textwidth]{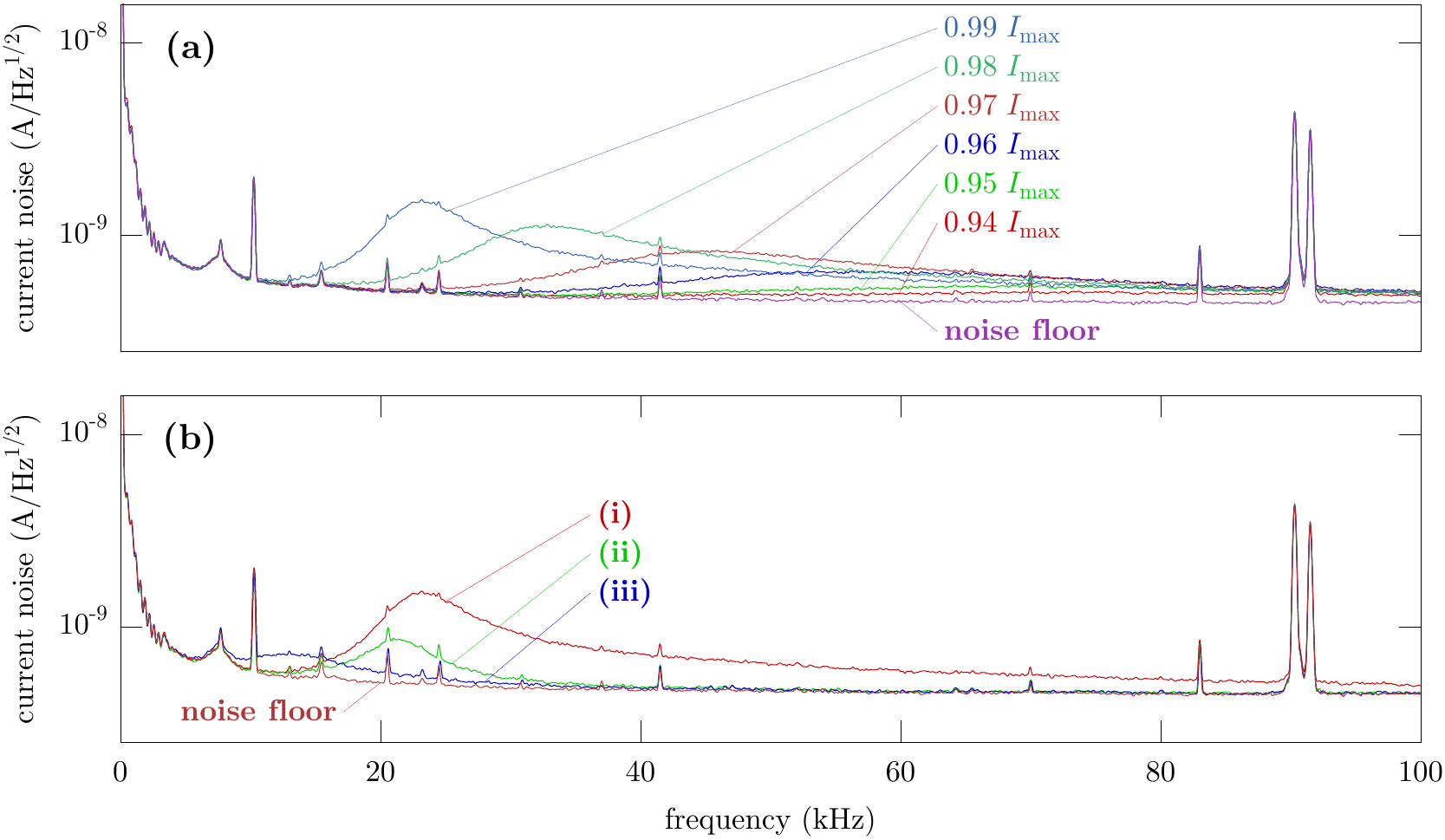}
\caption{Current-noise density of the Libbrecht--Hall style circuit.  (a), $C_1=10~\mu\mathrm{F}$, $C_2=10~\mu\mathrm{F}$, and $C_\mathrm{snub}=33~\mathrm{nF}$, and traces are plotted for several current set points (labeled as a fraction of $I_{\mathrm{max}}$).  The noise peak grows and moves from right to left as current is increased toward $I_{\mathrm{max}}$.  (b) shows variations of the circuit operating at 99\% of the maximum current.  The snubber resistor is $R_\mathrm{snub}=100~\Omega$, and capacitor values used are (i) $C_1=10~\mu\mathrm{F}$, $C_2=10~\mu\mathrm{F}$, $C_\mathrm{snub}=33~\mathrm{nF}$; (ii) $C_1=22~\mu\mathrm{F}$, $C_2=940~\mu\mathrm{F}$, $C_\mathrm{snub}=33~\mathrm{nF}$; and (iii) $C_1=22~\mu\mathrm{F}$, $C_2=940~\mu\mathrm{F}$, $C_\mathrm{snub}=100~\mathrm{nF}$.
\label{fig:noise}}
\end{figure*}

Fig.~\ref{fig:noise}(a) shows the behavior described in Sec.~\ref{sec:vreg}.  As the output current approaches $I_{\mathrm{\small{max}}}$, the feedback loop becomes less stable.  A noise peak appears around 100~kHz and shifts toward smaller frequencies.  Onset of this noise begins when the set current is within $\sim$10\% of $I_{\mathrm{\small{max}}}$.  Narrow peaks in the figure are due to pickup in our measurement section, since they are present in the baseline trace as well. 
For these data we use capacitor and snubber values specified in Ref.~\onlinecite{Troxel11} for the AD8671 and IRF9Z14 ($C_1=10~\mu\mathrm{F}$, $C_2=10~\mu\mathrm{F}$, $C_\mathrm{snub}=33~\mathrm{nF}$, and $R_\mathrm{snub}=100~\Omega$).  Note that the feedback circuit is not oscillating, and that the emerging noise peak is easy to overlook.

Characteristics of this noise peak are in part determined by the capacitors that help to stabilize the LM317 regulator. We found that increasing $C_1$ and $C_2$ reduced the amplitude and delayed the onset of the noise as current was increased, but capacitances larger than $C_1=22~\mu\mathrm{F}$ and $C_2=100~\mu\mathrm{F}$ did not further mitigate the problem.  We settled on a $22~\mu\mathrm{F}$ tantalum capacitor for $C_1$ and a pair of $470~\mu\mathrm{F}$ electrolytic capacitors for $C_2$.  Note that if little or no capacitance is used for $C_1$ and $C_2$, the circuit behaves quite poorly near $I_{\mathrm{\small{max}}}$: the noise peak is broader, an order of magnitude larger in amplitude, and is easily observed in the frequency-noise spectrum of an atomic absorption signal produced with a laser driven by the circuit. We also confirmed that removing the 5-$\Omega$, 44-$\mu\mathrm{F}$ low-pass filter at the LM317's output increases the noise, so these are important components to include.

Next we increase $C_\mathrm{snub}$, the snubber-network capacitor, and observe that this further reduces the noise peak and delays onset of its appearance as $I$ approaches $I_{\mathrm{\small{max}}}$.  To illustrate the improvements, we choose $I=0.99\,I_{\mathrm{\small{max}}}$ and plot data in Fig.~\ref{fig:noise}(b) for (i) original capacitor values from Ref.~\onlinecite{Troxel11}, (ii) altered LM317 bypass capacitors, and (iii) additionally altered snubber capacitor.  As seen in the plot, the capacitors in (iii) have significantly alleviated the noise when compared with (i).  Table~\ref{tab1} shows integrated current noise above the background for each of the three configurations at various output current levels.  Note that while the larger capacitors in (ii) and (iii) certainly stabilize the circuit as $I$ approaches $I_{\mathrm{\small{max}}}$, they also reduce the integrated noise when operating far away from $I_{\mathrm{\small{max}}}$.  Although integrated noise values are unavailable for comparison in Refs.~\onlinecite{Erickson08,Troxel11}, our values at 0.80\,$I_{\mathrm{\small{max}}}$ are comparable with those estimated in the original LH paper.\cite{Libbrecht93}

\begin{center}
\begin{table}[h!]
\begin{tabular}{>{\centering\arraybackslash}p{2cm}p{1.33cm}p{1.33cm}p{1.33cm}}
\multicolumn{1}{c}{Current                              } & \multicolumn{3}{c}{Integrated noise (nA)}\\
& \multicolumn{1}{c}{(i)} & \multicolumn{1}{c}{(ii)} & \multicolumn{1}{c}{(iii)}\\[2pt]
\hline
0.80 $I_{\mathrm{\small{max}}}$ & \multicolumn{1}{c}{35(6)} & \multicolumn{1}{c}{32(6)} & \multicolumn{1}{c}{28(7)} \\

0.94 $I_{\mathrm{\small{max}}}$ & \multicolumn{1}{c}{75(3)} & \multicolumn{1}{c}{46(4)} & \multicolumn{1}{c}{41(5)} \\

0.95 $I_{\mathrm{\small{max}}}$ & \multicolumn{1}{c}{90(2)} & \multicolumn{1}{c}{49(4)} & \multicolumn{1}{c}{43(5)} \\

0.96 $I_{\mathrm{\small{max}}}$ & \multicolumn{1}{c}{113(2)} & \multicolumn{1}{c}{45(5)} & \multicolumn{1}{c}{45(5)} \\

0.97 $I_{\mathrm{\small{max}}}$ & \multicolumn{1}{c}{138(1)} & \multicolumn{1}{c}{51(4)} & \multicolumn{1}{c}{49(4)} \\

0.98 $I_{\mathrm{\small{max}}}$ & \multicolumn{1}{c}{164(1)} & \multicolumn{1}{c}{60(3)} & \multicolumn{1}{c}{57(4)} \\

0.99 $I_{\mathrm{\small{max}}}$ & \multicolumn{1}{c}{175(1)} & \multicolumn{1}{c}{76(3)} & \multicolumn{1}{c}{64(3)} \\

\hline
 & & & \\
\end{tabular}
\caption{Current noise above background, integrated from 5~kHz to 200~kHz for various set points.  Capacitor values in (i), (ii), and (iii) are those listed in the caption of Fig.~\ref{fig:noise}(b).
\label{tab1}}
\end{table}
\end{center}

We also thought it interesting to compare the performance of the AD8671 to some other low-noise op-amps: the original Libbrecht--Hall LT1028, the LT1128, the AD797, and the ADA4898-1.  When using the capacitor values from circuit (iii), however, the AD797 and LT1028 are unstable and the snubber network must be adjusted to sufficiently load the output of the op-amp.\cite{IntersilApp}  We set $R_\mathrm{snub}$ to $50~\Omega$ and change $C_\mathrm{snub}$ to $200~\mathrm{nF}$, maintaining the same RC frequency.  Integrated noise values of the five op-amps are given in Table~\ref{tab2}.  
The performance of the AD8671 with this snubber network is better at low currents than in circuit (iii), but worse at high currents. The top overall performer is the AD797, which consistently has the lowest integrated current noise. Any of the five op-amps tested give low noise far away from $I_{\mathrm{\small{max}}}$, and it is possible that their performance could be improved with further changes to the snubber network. However, the
interaction of the capacitive load and the op-amp's output impedance create a frequency peaking unique to each op-amp/MOSFET combination, and there is no universal snubber-network solution. 

\begin{center}
\begin{table}[h!]
\begin{tabular}{>{\centering\arraybackslash}p{1.8cm}p{1.1cm}p{1.1cm}p{1.1cm}p{1.1cm}p{1.1cm}}
\multicolumn{1}{c}{Current} & \multicolumn{5}{c}{Integrated noise (nA)}\\
& \multicolumn{1}{c}{AD8671} & \multicolumn{1}{c}{ADA4898-1} & \multicolumn{1}{c}{AD797} & \multicolumn{1}{c}{LT1028} & \multicolumn{1}{c}{LT1128}\\[2pt]
\hline
0.80 $I_{\mathrm{\small{max}}}$ & \multicolumn{1}{c}{18(4)} & \multicolumn{1}{c}{12(6)} & \multicolumn{1}{c}{12(6)} & \multicolumn{1}{c}{17(4)} & \multicolumn{1}{c}{15(5)} \\

0.94 $I_{\mathrm{\small{max}}}$ & \multicolumn{1}{c}{40(2)} & \multicolumn{1}{c}{22(3)} & \multicolumn{1}{c}{15(5)} & \multicolumn{1}{c}{33(2)} & \multicolumn{1}{c}{34(2)} \\

0.95 $I_{\mathrm{\small{max}}}$ & \multicolumn{1}{c}{41(2)} & \multicolumn{1}{c}{26(3)} & \multicolumn{1}{c}{28(3)} & \multicolumn{1}{c}{48(2)} & \multicolumn{1}{c}{37(2)} \\

0.96 $I_{\mathrm{\small{max}}}$ & \multicolumn{1}{c}{45(2)} & \multicolumn{1}{c}{39(2)} & \multicolumn{1}{c}{36(2)} & \multicolumn{1}{c}{44(2)} & \multicolumn{1}{c}{46(2)} \\

0.97 $I_{\mathrm{\small{max}}}$ & \multicolumn{1}{c}{53(1)} & \multicolumn{1}{c}{56(1)} & \multicolumn{1}{c}{49(1)} & \multicolumn{1}{c}{55(1)} & \multicolumn{1}{c}{48(1)} \\

0.98 $I_{\mathrm{\small{max}}}$ & \multicolumn{1}{c}{63(1)} & \multicolumn{1}{c}{53(1)} & \multicolumn{1}{c}{53(1)} & \multicolumn{1}{c}{54(1)} & \multicolumn{1}{c}{61(1)} \\

0.99 $I_{\mathrm{\small{max}}}$ & \multicolumn{1}{c}{80(1)} & \multicolumn{1}{c}{57(1)} & \multicolumn{1}{c}{51(1)} & \multicolumn{1}{c}{57(1)} & \multicolumn{1}{c}{71(1)} \\

\hline
 & & & & &\\
\end{tabular}
\caption{Current noise above background, integrated from 5~kHz to 200~kHz for various set points. For these data we use $C_1=22~\mu\mathrm{F}$, $C_2=940~\mu\mathrm{F}$, $C_\mathrm{snub}=200~\mathrm{nF}$, and $R_\mathrm{snub}=50~\Omega$, and repeat the measurement for several low-noise integrated circuits.
\label{tab2}
}
\end{table}
\end{center}

\section{Conclusion\label{sec:conclusion}}

More than two decades after publication, the current driver design of Libbrecht and Hall remains a dependable low-noise solution for powering laser diodes. Users, however, should be aware of the effect on performance as the circuit's output current approaches the maximum set limit. The adjustments described in this work are simple to implement and improve the circuit's behavior.

\begin{acknowledgements}
This work was supported by NSF Grant No. PHY-1404455 and NSF Grant No. PHY-1505118.
\end{acknowledgements}


\begin{thebibliography}{9}
\expandafter\ifx\csname natexlab\endcsname\relax\def\natexlab#1{#1}\fi
\expandafter\ifx\csname bibnamefont\endcsname\relax
  \def\bibnamefont#1{#1}\fi
\expandafter\ifx\csname bibfnamefont\endcsname\relax
  \def\bibfnamefont#1{#1}\fi
\expandafter\ifx\csname citenamefont\endcsname\relax
  \def\citenamefont#1{#1}\fi
\expandafter\ifx\csname url\endcsname\relax
  \def\url#1{\texttt{#1}}\fi
\expandafter\ifx\csname urlprefix\endcsname\relax\def\urlprefix{URL }\fi
\providecommand{\bibinfo}[2]{#2}
\providecommand{\eprint}[2][]{\url{#2}}

\bibitem[{\citenamefont{Wieman and Hollberg}(1991)}]{Wieman91}
\bibinfo{author}{\bibfnamefont{C.~E.} \bibnamefont{Wieman}} \bibnamefont{and}
  \bibinfo{author}{\bibfnamefont{L.}~\bibnamefont{Hollberg}},
  \bibinfo{journal}{{Rev.\ Sci.\ Instrum.}} \textbf{\bibinfo{volume}{62}},
  \bibinfo{pages}{1} (\bibinfo{year}{1991}).

\bibitem[{\citenamefont{Libbrecht and Hall}(1993)}]{Libbrecht93}
\bibinfo{author}{\bibfnamefont{K.~G.} \bibnamefont{Libbrecht}}
  \bibnamefont{and} \bibinfo{author}{\bibfnamefont{J.~L.} \bibnamefont{Hall}},
  \bibinfo{journal}{{Rev.\ Sci.\ Instrum.}} \textbf{\bibinfo{volume}{64}},
  \bibinfo{pages}{2133} (\bibinfo{year}{1993}).

\bibitem[{\citenamefont{Erickson et~al.}(2008)\citenamefont{Erickson,
  Van~Zijll, Doermann, and Durfee}}]{Erickson08}
\bibinfo{author}{\bibfnamefont{C.~J.} \bibnamefont{Erickson}},
  \bibinfo{author}{\bibfnamefont{M.}~\bibnamefont{Van~Zijll}},
  \bibinfo{author}{\bibfnamefont{G.}~\bibnamefont{Doermann}}, \bibnamefont{and}
  \bibinfo{author}{\bibfnamefont{D.~S.} \bibnamefont{Durfee}},
  \bibinfo{journal}{Rev.\ Sci.\ Instrum.} \textbf{\bibinfo{volume}{79}},
  \bibinfo{eid}{073107} (\bibinfo{year}{2008}).

\bibitem[{\citenamefont{Troxel et~al.}(2011)\citenamefont{Troxel, Erickson, and
  Durfee}}]{Troxel11}
\bibinfo{author}{\bibfnamefont{D.~L.} \bibnamefont{Troxel}},
  \bibinfo{author}{\bibfnamefont{C.~J.} \bibnamefont{Erickson}},
  \bibnamefont{and} \bibinfo{author}{\bibfnamefont{D.~S.}
  \bibnamefont{Durfee}}, \bibinfo{journal}{Rev.\ Sci.\ Instrum.}
  \textbf{\bibinfo{volume}{82}}, \bibinfo{eid}{096101} (\bibinfo{year}{2011}).

\bibitem[{\citenamefont{Horowitz and Hill}(2016)}]{Horowitz}
\bibinfo{author}{\bibfnamefont{P.}~\bibnamefont{Horowitz}} \bibnamefont{and}
  \bibinfo{author}{\bibfnamefont{W.}~\bibnamefont{Hill}},
  \emph{\bibinfo{title}{The Art of Electronics}}
  (\bibinfo{publisher}{Cambridge}, \bibinfo{year}{2016}),
  \bibinfo{edition}{3rd} ed.

\bibitem[{\citenamefont{Dietz}()}]{Dietz}
\bibinfo{author}{\bibfnamefont{E.~H.} \bibnamefont{Dietz}},
  \bibinfo{note}{``Understanding and Reducing Noise Voltage on 3-Terminal
  Voltage Regulators,'' in Robert A. Pease, \textit{Troubleshooting Analog
  Circuits}, App.\ C, p.\ 191}.

\bibitem[{\citenamefont{Harvey and Siu}(1998)}]{IntersilApp}
\bibinfo{author}{\bibfnamefont{B.}~\bibnamefont{Harvey}} \bibnamefont{and}
  \bibinfo{author}{\bibfnamefont{C.}~\bibnamefont{Siu}},
  \emph{\bibinfo{title}{Driving reactive loads with high-frequency op-amps}},
  \bibinfo{howpublished}{Intersil Application Note AN1092}
  (\bibinfo{year}{1998}).

\bibitem[{\citenamefont{Blake}(2008)}]{MicrochipApp}
\bibinfo{author}{\bibfnamefont{K.}~\bibnamefont{Blake}},
  \emph{\bibinfo{title}{Driving capacitive loads with op-amps}},
  \bibinfo{howpublished}{Microchip Technology Inc. Application Note AN844}
  (\bibinfo{year}{2008}).

\bibitem[{\citenamefont{Sauer}(2004)}]{Sauer}
\bibinfo{author}{\bibfnamefont{J.~A.} \bibnamefont{Sauer}}, Ph.D. thesis,
  \bibinfo{school}{Georgia Institute of Technology} (\bibinfo{year}{2004}).

\end{thebibliography}
\end{document}